\documentclass[preprintnumbers, prd, floatfix,twocolumn,preprintnumbers, letterpaper, superscriptaddress,nofootinbib]{revtex4-1}
\usepackage{amsfonts}
\usepackage{amsmath}
\usepackage{amssymb,epsf}
\usepackage{latexsym}
\usepackage{graphicx,epsfig}
\usepackage{amssymb}
\usepackage{float}
\usepackage{subfigure}
\usepackage{epstopdf}
\usepackage[colorlinks=true,citecolor=blue,linkcolor=blue,urlcolor=black]{hyperref}
\usepackage{dcolumn}
\usepackage{psfrag}
\usepackage{wrapfig}
\usepackage{makeidx}
\usepackage{epsf}
\usepackage{color}
\usepackage{multirow}

\graphicspath{{Images/}}

\usepackage{tikz}
\definecolor{lime}{HTML}{A6CE39}
\newcommand{\orcidicon}{%
    \begin{tikzpicture}
    \draw[lime, fill=lime] (0,0)
        circle [radius=0.16]
        node[white] {{\fontfamily{qag}\selectfont \tiny ID}};
    \draw[white, fill=white] (-0.0625,0.095)
        circle [radius=0.007];
    \end{tikzpicture}   \hspace{-2mm}
}
\newcommand\orcidAhmad{{\href{https://orcid.org/0000-0002-7496-4270}{\orcidicon}}}
\newcommand\orcidMahdi{{\href{https://orcid.org/0000-0003-1196-9493}{\orcidicon}}}
\begin{document}
%========================================================
\title{Modified cosmology through Kaniadakis entropy}
%========================================================
\author{Mahdi Kord Zangeneh\orcidMahdi\!\!}
\email{mkzangeneh@scu.ac.ir}
\affiliation{Physics Department, Faculty of Science, Shahid Chamran University of Ahvaz,
Ahvaz 61357-43135, Iran}
%========================================================
\author{Ahmad Sheykhi\orcidAhmad\!\!}
\email{asheykhi@shirazu.ac.ir}
\affiliation{Department of Physics, College of Sciences, Shiraz University, Shiraz 71454, Iran}
\affiliation{Biruni Observatory, College of Sciences, Shiraz University, Shiraz 71454, Iran}
%========================================================
\begin{abstract}
We explore a cosmological model inspired by the modified
Kaniadakis entropy and disclose the influences of the modified
Friedmann equations on the evolution of the Universe. We find that
in modified Kaniadakis cosmology with only pressure-less matter, one
can reproduce the accelerated Universe without invoking any kind
of dark energy. We delved into the evolution of the Universe
during the radiation-dominated era as well. We also investigate
the behavior of the scale factor and the deceleration parameter
for a multiple-component Universe consisting of pressure-less
matter and a cosmological constant/dark energy. Interestingly
enough, the predicted age of the Universe in the modified
Kaniadakis cosmology becomes larger compared to the standard
cosmology which may alleviate the age problem. Furthermore, The
findings reveal that in the modified Kaniadakis cosmology, the
transition from a decelerated phase to an accelerated Universe
occurs at higher redshifts compared to the standard cosmology.
These results shed light on the potential implications of
incorporating Kaniadakis entropy into cosmological models and
provide valuable insights into the behavior of the Universe in
different cosmological scenarios. Moreover, they emphasize the
crucial role of modifications to the geometry component and the
significance of such modifications in understanding the dynamics
of the Universe.
\end{abstract}
%========================================================
\maketitle
%========================================================

\section{Introduction\label{Intro}}

The Kaniadakis entropy is one-parameter entropy that generalizes the
Boltzmann-Gibbs-Shannon entropy. It is derived from a relativistic
statistical theory and preserves the basic features of standard statistical
theory \cite{Kan1,Kan2}. The Kaniadakis entropy is given by a general
expression involving the Kaniadakis parameter, denoted as $K$, which
measures the deviation from standard statistical mechanics.
%In this generalized statistical theory, the distribution function is given by an exponential function that depends on the Kaniadakis parameter.% The chemical potential can be determined by normalization.

In terms of the probability in which the system to be in a specific
microstate, $P_i$, the Kaniadakis entropy can be expressed as $S_{K}
=-k_{_B}\sum^{W}_{i=1}\left(P^{1+K}_{i}-P^{1-K}_{i}\right)/2K$, where $k_{_B}
$ is Boltzmann constant and $W$ is the total number of the system
configurations \cite%
{Abreu:2016avj,Abreu:2017fhw,Abreu:2017hiy,Abreu:2018mti,Yang:2020ria,Abreu:2021avp}%
. Hereafter we set $k_{_B}=c=\hbar=1$.

The Kaniadakis entropy can also be applied to black hole thermodynamics. In
Einstein gravity, the entropy of a black hole follows the Bekenstein-Hawking
entropy, which states that the entropy of the black hole is proportional to
the area of the horizon and is given by $S_{BH}=A/\left( 4 G \right)$ where $%
G$ is the Newtonian gravitational constant. By assuming $P_i=1/W$ and using
the Boltzmann-Gibbs entropy, we can relate the probability to the
Bekenstein-Hawking entropy as $P_i=\exp{(-S_{BH})}$ and arrive at $S_{K} =
K^{-1}\sinh{(K S_{BH})}$ \cite{Mor}. When the Kaniadakis parameter $K$
approaches zero, the standard Bekenstein-Hawking entropy is recovered. Since
the deviation from the standard entropy is expected to be small, we can
expand the Kaniadakis entropy \cite{Sheykhi:2023aqa}
\begin{equation}  \label{kentropy2}
S_{K}=S_{BH}+ \frac{K^2}{6} S_{BH}^3+ {\mathcal{O}}(K^4).
\end{equation}
This correction term represents the leading order Kaniadakis correction to
the black hole entropy.

In \cite{Sheykhi:2023aqa}, the Kaniadakis entropy has been used to find
modified Friedmann equations. Already, the Kaniadakis entropy was also used
to find modified Friedmann equations, with the difference being that this
modification was appeared as new extra terms that constitute an effective
dark energy sector \cite{Lym}. The cosmological implications of modified
Friedmann equations given in \cite{Lym} were also explored in \cite%
{Luci,Her,Dre,P:2022amn}. Since entropy has close relationship with
spacetime geometry, it is expected that any modification to the entropy
expression alter the geometric part of the equations. This approach has been
used in \cite{Sheykhi:2023aqa}, and therefore, here, we will use the
modified equations of this recent paper to study Kaniadakis cosmology.

The modified Friedmann equations due to generalized Tsallis and Barrow
entropies have also been derived in \cite{Sheykhi:2018dpn} and \cite%
{Sheykhi:2021fwh}, respectively. Based on the modified equations,
cosmological implication were explored for both cases \cite%
{Lymperis:2018iuz,Sheykhi:2022jqq}. The cosmological model based on Tsallis
entropy follows a usual thermal history with different eras of matter and
dark energy \cite{Lymperis:2018iuz}. The dark energy equation of state
parameter can also take on different values, resulting in quintessence-like,
phantom-like, or phantom-divide crossing behavior during its evolution \cite%
{Lymperis:2018iuz}. In Barrow cosmology, the transition from deceleration to
acceleration phase occurs at lower redshifts in comparison with standard
cosmology \cite{Sheykhi:2022jqq}. The estimated age of the Universe in this
model is also smaller than in standard cosmology \cite{Sheykhi:2022jqq}.
Other aspects, such as observational constraints and growth of
perturbations, have been investigated in the contexts of both Tsallis and
Barrow cosmologies as well \cite%
{Asghari:2021lzu,Asghari:2021bqa,Sheykhi:2022gzb}. The objective of present
study is to examine the cosmological consequences that arise from the
modified Friedmann equations when the Kaniadakis entropy is used as the
entropy associated with the apparent horizon \cite{Sheykhi:2023aqa}.

This paper is structured as follows. In the next section we review the
derivation of the modified Friedmann equations based on Kaniadakis entropy.
In section \ref{Cosmology}, we investigate the cosmological consequences of the modified
Friedmann equations during the history of the Universe. We finish with summary and
closing remarks in the last section.

%%%%%%%%%%%%%%%%%%%%%%%%%%%%%%%%%%%%%%%%%%%%%%%%%%%%%%%%%%%%%%%%%%%%%%%%%%%%%%%%%%%%%%%%%%%%

\section{Modified Friedmann equations inspired by Kaniadakis entropy\label%
{first}}

In this section, we review the derivation of the modified Friedmann
equations based on Kaniadakis entropy, utilizing the gravity-thermodynamics
conjecture. Detailed calculations can be found in \cite{Sheykhi:2023aqa}.

In the background of the FRW Universe, the metric's line element is
expressed as
\begin{equation}
ds^{2}=-dt^{2}+a^{2}(t)\left( \frac{dr^{2}}{1-kr^{2}}+r^{2}(d\theta
^{2}+\sin ^{2}\theta d\phi ^{2})\right) ,
\end{equation}%
where $a(t)$ is scale factor of the Universe, $k=0,\pm 1$ represents the
curvature parameter, and $(t,r,\theta ,\phi )$ are the co-moving
coordinates. We make the assumption that $a_{0}=a(t=t_{0})=1$, representing
the scale factor at the present time. Taking the apparent horizon as the
boundary of the Universe, the temperature associated with the horizon is
determined by \cite{Cai1,Shey1,Shey2,Shey3}
\begin{equation}
T_{h}=-\frac{1}{2\pi \tilde{r}_{A}}\left( 1-\frac{\dot{\tilde{r}}_{A}}{2H%
\tilde{r}_{A}}\right) ,  \label{T}
\end{equation}%
where $\tilde{r}_{A}={1}/\sqrt{H^{2}+k/a^{2}}$ is the apparent horizon
radius \cite{Sheyem}. From a thermodynamic perspective, the apparent horizon
can be considered a suitable horizon that adheres to the first and second
laws of thermodynamics \cite{wang1,wang2,Cai2, Cai3,SheyCQ,sheyECFE}. In
addition, we make the assumption that the energy-momentum tensor of the
Universe is given by $T_{\mu \nu }=(\rho +p)u_{\mu }u_{\nu }+pg_{\mu \nu }$,
where $\rho $ represents the energy density and $p$ represents the pressure.
This energy-momentum tensor is conserved, leading to the continuity equation
$\nabla _{\mu }T^{\mu \nu }=0$, which can be expressed as $\dot{\rho}%
+3H(\rho +p)=0$, where $H=\dot{a}/a$ represents the Hubble parameter. The
work density associated with the expansion of the Universe is denoted as $%
W=(\rho -p)/2$ \cite{Hay2}. In order to apply the gravity-thermodynamics
conjecture, we propose that the first law of thermodynamics on the apparent
horizon is satisfied:
\begin{equation}
dE=T_{h}dS_{h}+WdV.  \label{FL}
\end{equation}%
Let us consider the total energy of the Universe enclosed by the apparent
horizon $E=\rho V$, and $T_{h}$ and $S_{h}$ are the temperature and entropy
associated with the apparent horizon, respectively. Here, $V=\frac{4\pi }{3}%
\tilde{r}_{A}^{3}$ denotes the volume encompassed by a three-dimensional
sphere with an apparent horizon area of $A=4\pi \tilde{r}_{A}^{2}$.

By taking the differential form of the total matter and energy, we obtain $%
dE=4\pi \tilde{r}_{A}^{2}\rho d\tilde{r}_{A}+\frac{4\pi }{3}\tilde{r}_{A}^{3}%
\dot{\rho}dt$. Combining this with the conservation equation, we can derive
the following expression:
\begin{equation}
dE=4\pi \tilde{r}_{A}^{2}\rho d\tilde{r}_{A}-4\pi H\tilde{r}_{A}^{3}(\rho
+p)dt.  \label{dE2}
\end{equation}%
Differentiating the Kaniadakis entropy, $S_{h}=\mathcal{S}+\frac{K^{2}}{6}%
\mathcal{S}^{3}$, yields
\begin{equation}
dS_{h}=d\mathcal{S}+\frac{K^{2}}{2}\mathcal{S}^{2}d\mathcal{S},  \label{dS}
\end{equation}%
where
\begin{equation*}
\mathcal{S}=\frac{A}{4G}=\frac{\pi \tilde{r}_{A}^{2}}{G}, d\mathcal{S}=\frac{%
2\pi \tilde{r}_{A}\dot{\tilde{r}}_{A}}{G}dt.
\end{equation*}%
Combining Eqs. (\ref{T}), (\ref{dE2}), and (\ref{dS}) with the first law of
thermodynamics (\ref{FL}), and using the continuity relation, after
performing some algebraic calculations, we can arrive at the following
expression
\begin{equation}
-\frac{2d\tilde{r}_{A}}{\tilde{r}_{A}^{3}}\left( 1+\alpha \tilde{r}%
_{A}^{4}\right) =\frac{8\pi G}{3}d\rho,  \label{Fried2}
\end{equation}
where we have defined the dimensionless parameter
\begin{equation*}
\alpha \equiv \frac{K^{2}\pi ^{2}}{2G^{2}}.
\end{equation*}%
After integration, we obtain the modified first Friedmann equation in
Kaniadakis cosmology \cite{Sheykhi:2023aqa}
\begin{equation}
H^{2}+\frac{k}{a^{2}}-\alpha \left( H^{2}+\frac{k}{a^{2}}\right) ^{-1}=\frac{%
8\pi G}{3}(\rho +\rho _{\Lambda }),  \label{Fried4}
\end{equation}%
where $\rho _{\Lambda }=\Lambda /(8\pi G)$ and $\Lambda $ is a constant of
integration which can be interpreted as the cosmological constant.
Therefore, by utilizing the first law of thermodynamics and assuming that
the entropy associated with the apparent horizon takes the Kaniadakis form,
we derive the modified Friedmann equation. It is worth noting that when $%
\alpha\rightarrow 0$, the resulting Friedmann equation coincides with one in
standard cosmology.

The second Friedmann equation can be derived by combining the continuity
equation with the first Friedmann equation (\ref{Fried4}). By performing the
necessary calculations, we can obtain the desired result \cite%
{Sheykhi:2023aqa}:

\begin{equation}
\left( \dot{H}-\frac{k}{a^{2}}\right) \left[ 1+\alpha \left( H^{2}+\frac{k}{%
a^{2}}\right) ^{-2}\right] =-4\pi G(\rho +p).  \label{2Fried3}
\end{equation}%
The second modified Friedmann equation, which describes the evolution of the
Universe according to Kaniadakis entropy, is also useful in exploring the
evolution of the Universe. It is worth noting that in the limit where $\alpha
$ approaches zero, Eq. (\ref{2Fried3}) simplifies to the conventional second
Friedmann equation found in standard cosmology.

With the modified Friedmann equations (\ref{Fried4}) and (\ref{2Fried3}) at
our disposal, we proceed to explore the cosmological consequences of this
model in the following section.
%%%%%%%%%%%%%%%%%%%%%%%%%%%%%%%%%%%%%%%%%%%%%%%%%%%%%%%%%%%%%%%%%%%%%%%%%%%%%%%%%%%%%%%%%%%%%%%%%%%%%

\section{Modified Kaniadakis Cosmology \label{Cosmology}}

In the upcoming section, we will explore the cosmological implications that
arise from the modified Friedmann equations, as presented in Eqs. (\ref%
{Fried4}) and (\ref{2Fried3}). To simplify our analysis, we will primarily
focus on the scenario of a flat Universe, denoted by $k=0$, although it is
worth noting that the study can be extended to include cases where $k=\pm 1$%
. In the context of a flat Universe, the first modified Friedmann equation
within Kaniadakis cosmology (\ref{Fried4}) reduces to
\begin{equation}
H^{2}-\alpha H^{-2}=\frac{8\pi G}{3}\left( \rho +\rho _{\Lambda }\right) .
\end{equation}%
By integrating the continuity equation $\dot{\rho}+3H(\rho +p)=0$ for matter
with an equation of state $p=\omega\rho$, we obtain $\rho(t)=\rho_0a^{-3(1+%
\omega)}$, where $\rho_0$ represents the matter energy density at the
present time, with $a_0=1$. Using this relation for energy density of matter
and the critical density of the Universe at the present time $%
\rho_{c,0}=3H_0^2/(8\pi G)$, where $H_0$ is the value of the Hubble
parameter at the present time, we can rewrite the above equation as
\begin{equation}
\left( \frac{H}{H_{0}}\right) ^{2}-\beta \left( \frac{H}{H_{0}}\right)
^{-2}=\Omega ^{0}a^{-3\left( 1+\omega \right) }+\Omega _{\Lambda }^{0}.
\label{2Fried4}
\end{equation}
where $\Omega ^{0}=\rho _{0}/\rho _{c,0}$ and $\Omega _{\Lambda
}^{0}=\rho _{\Lambda }/\rho _{c,0}$ are density parameters of
matter (with equation of state $p=\omega \rho $) and cosmological
constant/dark energy, respectively and $\beta =\alpha H_{0}^{-4}$
is a dimensionless parameter.

In the subsequent analysis, we will utilize Eq. (\ref{2Fried4}) to examine
the cosmological implications arising from the Kaniadakis cosmology. This
equation will allow us to explore the effects and consequences of the
modified framework on various aspects of the Universe.

\subsection{Single-component flat Universes}

\subsubsection*{Matter-dominated era}

To begin our analysis, we will consider a scenario where the cosmological
constant is zero ($\Lambda =0$) and the Universe is dominated by
pressure-less matter ($\omega =0$). In this case, the modified Friedmann
equation (\ref{Fried4}) reduces to

\begin{equation}
\left( \frac{H}{H_{0}}\right) ^{2}-\beta \left( \frac{H}{H_{0}}\right)
^{-2}=\Omega _{m}^{0}a^{-3}.  \label{3Fried4}
\end{equation}%
In this particular scenario, we find that at the present time,
when $H=H_{0}$ and $a=a_{0}=1$, the sum of the matter density
parameter $\Omega _{m}^{0}$ and the dimensionless parameter
$\beta$ that relates to the Kaniadakis parameter is equal to $1$,
i.e., $\Omega _{m}^{0}+\beta =1$. This indicates the presence of
an additional term arising from the Kaniadakis entropy correction
to the Friedmann equations which can push our Universe to
accelerate without invoking any kind of dark energy. We can rewrite
the modified Friedmann equation as $\left( H/H_{0}\right)
^{4}-\Omega _{m}^{0}a^{-3}\left( H/H_{0}\right) ^{2}-\beta =0$. By
solving this equation, we can obtain the following solution

\begin{equation}
\left( \frac{H}{H_{0}}\right) ^{2}=\frac{\Omega _{m}^{0}a^{-3}+\sqrt{\left(
\Omega _{m}^{0}a^{-3}\right) ^{2}+4\beta }}{2}.
\end{equation}%
For small values of $\beta$, the solution can be approximated as $\left(
H/H_{0}\right) ^{2}=\Omega _{m}^{0}a^{-3}+\beta a^{3}/\Omega _{m}^{0}+%
\mathrm{O}\left(\beta ^{2}\right)$. Hence, in this limit, we obtain

\begin{gather*}
\frac{da}{a\sqrt{\Omega _{m}^{0}/a^{3}+\beta a^{3}/\Omega _{m}^{0}}}=H_{0}dt,
\\
\Rightarrow \left( \sqrt{\frac{a}{\Omega _{m}^{0}}}-\beta \frac{a^{13/2}}{%
2\left( \Omega _{m}^{0}\right) ^{5/2}}\right) da=H_{0}dt,
\end{gather*}%
where we have used $H=(da/dt)/a$. By integrating the aforementioned
relation, we arrive at

\begin{equation}
\frac{2a^{3/2}}{3\sqrt{\Omega _{m}^{0}}}-\beta \frac{a^{15/2}}{15\left(
\Omega _{m}^{0}\right) ^{5/2}}=H_{0}t,  \label{atmd}
\end{equation}
where we have set the integration constant equal to zero, for simplicity.
Indeed this is the case if we assume $a \rightarrow 0$ as $t \rightarrow 0$.
The above equation establishes the connection between time and the scale
factor. By utilizing this equation, we can study the behavior of the scale
factor as a function of time for various sets of parameters.

\begin{figure}[t]
\includegraphics[scale=0.8]{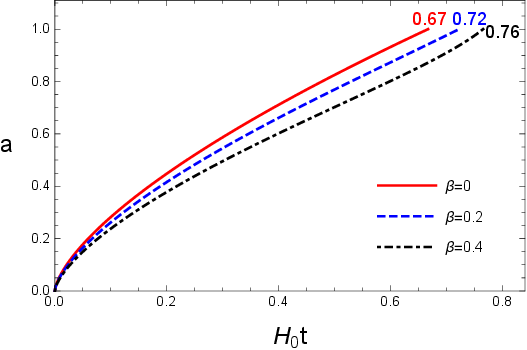}
\caption{The behavior of scale factor $a(t)$ vs $H_{0}t$ in modified
Kaniadakis cosmology for matter-dominated era with different values of $%
\protect\beta $. Note that $\Omega _{m}^{0}=1-\protect\beta $.}
\label{fig1}
\end{figure}
%%%%%%%%%%%%%%%%%%%%%%%%%%%%%%%%%%%%%%%%%%%%%%%%%%%
\begin{table}[t]
\begin{center}
\begin{tabular}{c|c|c|c|}
\hline
\multicolumn{1}{|c|}{$\beta$} & $0$ & $0.2$ & $0.4$   \\ \hline
\multicolumn{1}{|c|}{$H_0 t_{0}$} & $0.67$ & $0.72$ & $0.76$   \\ \hline
\end{tabular}%
\end{center}
\caption{The values of $H_0 t_0$ in modified Kaniadakis cosmology for a
matter-dominated Universe and different values of $\protect\beta=\protect%
\alpha H_0^{-4}$.}
\label{table1}
\end{table}

In Fig. \ref{fig1}, we observe the behavior of the scale factor, $a(t)$, for
different values of $\beta$. It is evident that for larger values of $\beta$%
, the model predicts a greater age for the Universe. In table \ref{table1},
we have given the age of the Universe for different values of $\beta$, which
indicates that in Kaniadakis cosmology, with only pressure-less matter, the
age problem can be alleviated. It was already argued that in Tsallis
cosmology, the age problem can also be resolved for specific values of the
Tsallis parameter \cite{SheTs}. However, it is important to note that not
all modified cosmological models based on correction to entropy, provide a
solution to the age problem. For instance, in Barrow cosmology, the age
problem cannot be resolved \cite{Sheykhi:2022jqq}.

Additionally, we observe that at each given time, the scale factor decreases
as $\beta $ increases. Consequently, in the modified Kaniadakis cosmology,
the radius of the Universe decreases in comparison to the standard
cosmology.

\begin{figure}[t]
\includegraphics[scale=0.8]{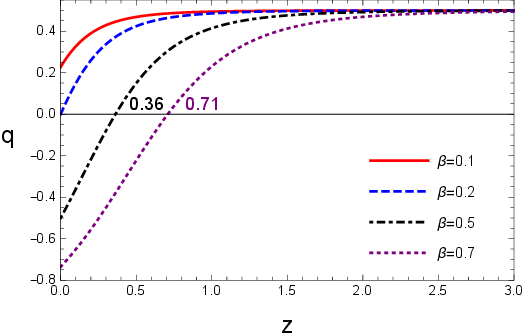}
\caption{The behavior of deceleration parameter $q$ vs redshift $z$ in
modified Kaniadakis cosmology for matter-dominated era with different values
of $\protect\beta $. Note that $\Omega _{m}^{0}=1-\protect\beta $.}
\label{fig2}
\end{figure}

The deceleration parameter, defined as $q=-1-\dot{H}/{H^{2}}$, for
a Kaniadakis Universe filled with pressure-less matter can be
determined using the equation:

\begin{equation}
q=-1+\frac{3}{2}\frac{\left( \Omega _{m}^{0}a^{-3}+\sqrt{(\Omega
_{m}^{0}a^{-3})^{2}+4\beta }\right) ^{2}-4\beta }{\left( \Omega
_{m}^{0}a^{-3}+\sqrt{(\Omega _{m}^{0}a^{-3})^{2}+4\beta }\right) ^{2}+4\beta}.
\end{equation}
This expression is derived using equations (\ref{2Fried3}) and
(\ref{3Fried4}). In Fig. \ref{fig2}, we present the behavior of
the deceleration parameter as a function of redshift $z=1/a-1$.
The plot demonstrates that for larger values of $\beta$,
specifically larger than a critical value near $\beta =0.2$, the
Universe experiences late-time accelerated expansion with only
pressure-less matter. This finding aligns with the results
obtained in \cite{salehi}, where the author establishes that by
setting cosmographic parameters, a Kaniadakis Universe with nearly
zero cosmological constant filled solely with pressure-less matter
can undergo late-time accelerated expansion. These results
highlight the impact of the modification introduced by Kaniadakis
entropy in the geometric component of field equations, which can
lead to accelerated expansion even in the absence of dark energy.
A similar outcome was previously reported in
\cite{Dehghani:2004cf}, where the inclusion of the Gauss-Bonnet
correction term in the geometry component resulted in the
accelerated expansion of the Universe.

These findings highlight the potential implications of Kaniadakis entropy in
cosmology, specifically in addressing the age problem and influencing the
evolution scenario of the Universe.

%%%%%%%%%%%%%%%%%%%%%%%%%%%%%%%%%%%%%%%%%%%%%%%%%%%%%%%%%%%%%%%%%%%%%%%%%%%%%%%

\subsubsection*{Radiation-dominated era}

\begin{figure}[t]
\includegraphics[scale=0.8]{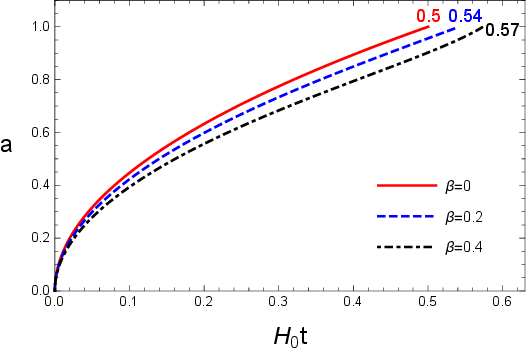}
\caption{The behavior of scale factor $a(t)$ vs $H_{0}t$ in modified
Kaniadakis cosmology for radiation-dominated era with different values of $%
\protect\beta $. Note that $\Omega _{r}^{0}=1-\protect\beta $.}
\label{fig3}
\end{figure}

Now let us consider a Universe that is filled with radiation. This scenario
is only relevant during the early stages of the Universe when radiation
dominated. As our Universe expands, the momentum of freely moving particles
decreases according to the relationship $P(t)\sim 1/a(t)$, where $a(t)$
represents the scale factor. This implies that the random velocities of
particles we observe today must have been much larger in the past when the
scale factor was significantly smaller than its present value \cite{GRbook}.
Consequently, the pressure-less approximation breaks down in the early
Universe. Our objective here is to determine the evolution of the Universe
within the framework of Kaniadakis cosmology, where the energy content of
the Universe consists of highly relativistic gas (radiation) with a pressure
given by $p=\rho/3$. In this particular case, with $\omega =1/3$, it is easy
to show that
\begin{equation}
\left( \frac{H}{H_{0}}\right) ^{2}-\beta \left( \frac{H}{H_{0}}\right)
^{-2}=\Omega _{r}^{0}a^{-4}.
\end{equation}%
At the present time, we have $\Omega _{r}^{0}+\beta =1$. We can rewrite the
equation as $\left( H/H_{0}\right) ^{4}-\Omega _{r}^{0}a^{-4}\left(
H/H_{0}\right) ^{2}-\beta =0$, which admits the following solution
\begin{equation}
\left( \frac{H}{H_{0}}\right) ^{2}=\frac{\Omega _{r}^{0}a^{-4}+\sqrt{\left(
\Omega _{r}^{0}a^{-4}\right) ^{2}+4\beta }}{2}.
\end{equation}
For small values of $\beta$, the solution simplifies to $\left(
H/H_{0}\right) ^{2}=\Omega _{r}^{0}a^{-4}+\beta a^{4}/\Omega _{r}^{0}+%
\mathrm{O}\left( \beta ^{2}\right)$. In this limit, we have
\begin{gather*}
\frac{da}{a\sqrt{\Omega _{r}^{0}/a^{4}+\beta a^{4}/\Omega _{r}^{0}}}=H_{0}dt,
\\
\Rightarrow \left( \frac{a}{\sqrt{\Omega _{r}^{0}}}-\beta \frac{a^{9}}{%
2\left( \Omega _{r}^{0}\right) ^{5/2}}\right) da=H_{0}dt.
\end{gather*}
By integrating the above relation, we arrive at
\begin{equation}
\frac{a^{2}}{2\sqrt{\Omega _{r}^{0}}}-\beta \frac{a^{10}}{20\left( \Omega
_{r}^{0}\right) ^{5/2}}=H_{0}t.
\end{equation}
Figure \ref{fig3} illustrates the behavior of the scale factor $a(t)$ for
different values of $\beta$ for a radiation dominated Universe in Kaniadakis
cosmology. We observe that as $\beta$ increases, the predicted age of the
Universe increases too. In table \ref{table2}, we present the age of the
Universe in Kaniadakis cosmology for radiation-dominated Universe and for
different values of $\beta$. It is seen that in the modified cosmology based
on Kaniadakis entropy, the age problem in standard cosmology can be
mitigated. Additionally, it is worth noting that at each moment in time, the
scale factor decreases as $\beta$ increases. Consequently, in the modified
Kaniadakis cosmology, the radius of the Universe is smaller compared to
standard cosmology.

\begin{table}[t]
\begin{center}
\begin{tabular}{c|c|c|c|}
\hline
\multicolumn{1}{|c|}{$\beta$} & $0$ & $0.2$ & $0.4$   \\ \hline
\multicolumn{1}{|c|}{$H_0 t_{0}$} & $0.50$ & $0.54$ & $0.57$  \\ \hline
\end{tabular}%
\end{center}
\caption{The values of $H_0 t_0$ in modified Kaniadakis cosmology for a
radiation-dominated Universe and different values of $\protect\beta=\protect%
\alpha H_0^{-4}$.}
\label{table2}
\end{table}
%%%%%%%%%%%%%%%%%%%%%%%%%%%%%%%%%%%%%%%%%%%%%%%%%%%%%%%%%%%%%%%%%%%%%%%%%%%%%%%%%%%%%%%%%%%

\subsection{Multiple-component flat Universe}

In a flat Universe with multiple components, including pressure-less matter,
radiation and cosmological constant/dark energy, the first modified
Friedmann equation (\ref{2Fried4}) can be written as
\begin{equation*}
\left( \frac{H}{H_{0}}\right) ^{2}-\beta \left( \frac{H}{H_{0}}\right)
^{-2}=\Omega _{m}^{0}a^{-3}+\Omega _{r}^{0}a^{-4}+\Omega _{\Lambda }^{0}.
\end{equation*}
Since the contribution of radiation density is negligible compared to the
other components \cite{Cosmobooks}, we can ignore it. Therefore, at the
present time, we have $\Omega _{m}^{0}+\Omega _{\Lambda }^{0}+\beta =1$.
Following the procedure of the previous subsections, we find

\begin{figure}[t]
\includegraphics[scale=0.8]{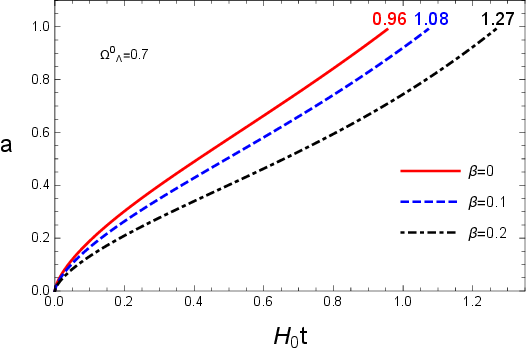}
\caption{The behavior of scale factor $a(t)$ vs $H_{0}t$ in modified
Kaniadakis cosmology for a multiple-components Universe with different
values of $\protect\beta $. Note that $\Omega _{m}^{0}=1-\Omega _{\Lambda
}^{0}-\protect\beta $.}
\label{fig4}
\end{figure}

%\begin{eqnarray*}
%\left( \frac{H}{H_{0}}\right) ^{4}-\left( \frac{H}{H_{0}}\right) ^{2}\left(
%\Omega _{m}^{0}a^{-3}+\Omega _{\Lambda }^{0}\right) -\beta  &=&0 \\
%\left( \frac{H}{H_{0}}\right) ^{2} &=&\frac{\left( \Omega
%_{m}^{0}a^{-3}+\Omega _{\Lambda }^{0}\right) +\sqrt{(\Omega
%_{m}^{0}a^{-3}+\Omega _{\Lambda }^{0})^{2}+4\beta }}{2} \\
%\int_{0}^{a}\left( \sqrt{\frac{a}{\left( \Omega _{m}^{0}+\Omega _{\Lambda
%}^{0}a^{3}\right) }}-\beta \frac{a^{13/2}}{2\left( \Omega _{m}^{0}+\Omega
%_{\Lambda }^{0}a^{3}\right) ^{5/2}}\right) da &=&H_{0}t
%\end{eqnarray*}%
\begin{eqnarray}
&& H_{0}t =\frac{2\sinh ^{-1}\left( \sqrt{\frac{\Omega _{\Lambda }^{0}}{%
\Omega _{m}^{0}}}a^{3/2}\right) }{3\sqrt{\Omega _{\Lambda }^{0}}}+\frac{%
\beta }{3\Omega _{\Lambda }^{05/2}}  \notag \\
&&\times \left[ \frac{a^{3/2}\sqrt{\Omega _{\Lambda }^{0}}\left( 3\Omega
_{m}^{0}+4a^{3}\Omega _{\Lambda }^{0}\right) }{3\left( \Omega
_{m}^{0}+a^{3}\Omega _{\Lambda }^{0}\right) ^{3/2}}-\sinh ^{-1}\left( \sqrt{%
\frac{\Omega _{\Lambda }^{0}}{\Omega _{m}^{0}}}a^{3/2}\right) \right].
\notag \\
&&
\end{eqnarray}
By utilizing the aforementioned equation, we can examine the behavior of the
scale factor $a(t)$ in a flat Universe with matter and cosmological
constant. Figure \ref{fig4} shows the behavior of the scale factor $a(t)$
for different values of $\beta$. It is evident that as $\beta$ increases,
the model predicts a greater age for the Universe. In particular, for $\beta
= 0$, $\beta = 0.1$ and $\beta = 0.2$ the corresponding ages are,
respectively, $t_{0} = 0.96H_{0}^{-1}$, $t_{0} = 1.08H_{0}^{-1}$ and $t_{0}
= 1.27H_{0}^{-1}$. This indicates that in a modified Kaniadakis cosmology
filled with pressure-less matter and dark energy/cosmological constant, the
age of the Universe increases compared to standard cosmology. Moreover, it
can be observed that at each point in time, the scale factor decreases with
increasing $\beta$. Consequently, in the modified Kaniadakis cosmology, the
radius of the Universe decreases compared to the standard cosmology. This
suggests that the modified cosmology leads to a contraction of the Universe,
potentially resulting in different physical implications and dynamics.
Overall, the behavior of $a(t)$ in Fig. \ref{fig4} highlights the potential
of the modified Kaniadakis cosmology to address the age problem and its
impact on the size and evolution of the Universe.

\begin{figure}[t]
\includegraphics[scale=0.8]{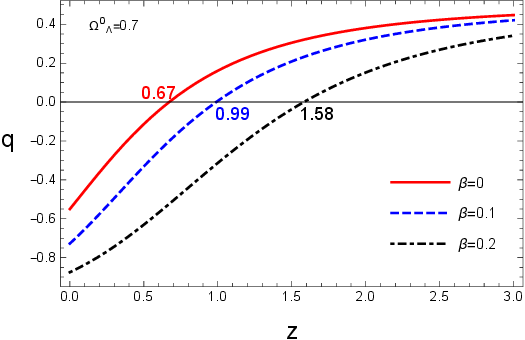}
\caption{The behavior of deceleration parameter $q$ vs redshift $z$ in
modified Kaniadakis cosmology for multiple-components Universe with
different values of $\protect\beta $. Note that $\Omega _{m}^{0}=1-\Omega
_{\Lambda }^{0}-\protect\beta $.}
\label{fig5}
\end{figure}

We can also calculate the deceleration parameter, defined as $q=-1-\dot{H}/{%
H^{2}}$, using equations (\ref{2Fried3}) and (\ref{2Fried4}). By performing
the necessary calculations, we find that

\begin{equation}
q=-1+\frac{3}{2}\frac{X-4\beta }{X+4\beta }\frac{\Omega _{m}^{0}a^{-3}}{%
\Omega _{\Lambda }^{0}+\Omega _{m}^{0}a^{-3}}.
\end{equation}%
where

\begin{equation*}
X=\left( \Omega _{m}^{0}a^{-3}+\Omega _{\Lambda }^{0}+\sqrt{\left( \Omega
_{m}^{0}a^{-3}+\Omega _{\Lambda }^{0}\right) ^{2}+4\beta }\right) ^{2}.
\end{equation*}
%\begin{eqnarray}
%q~&& =-1+\frac{3\Omega _{m}^{0}}{2a^{3}}  \notag \\
%\times && \left[ \frac{\left[ \left( \Omega _{m}^{0}a^{-3}+\Omega _{\Lambda
%}^{0}\right) +\sqrt{(\Omega _{m}^{0}a^{-3}+\Omega _{\Lambda
%}^{0})^{2}+4\beta }\right] ^{2}+4\beta }{2\left( \Omega
%_{m}^{0}a^{-3}+\Omega _{\Lambda }^{0}\right) +\sqrt{(\Omega
%_{m}^{0}a^{-3}+\Omega _{\Lambda }^{0})^{2}+4\beta }}\right] ^{-1}.  \notag \\
%&&
%\end{eqnarray}%
In the context of the modified Kaniadakis cosmology, the behavior of the
deceleration parameter $q$ as a function of redshift $z$ provides valuable
insights. By expressing the scale factor as $a=(1+z)^{-1}$, we can examine
the transition from a decelerating phase ($q>0$) to an accelerating phase ($%
q<0$). Figure \ref{fig5} visually represents the relationship between $q$
and $z$, and it reveals an interesting trend. As the parameter $\beta$
increases, the transition from deceleration to acceleration occurs at higher
redshifts. This observation suggests that in the modified Kaniadakis
cosmology, the onset of cosmic acceleration happens earlier compared to the
standard cosmology. This finding has significant implications for our
understanding of the evolution of the Universe. The modified Kaniadakis
cosmology indicates that the Universe transitions from deceleration to
acceleration at an earlier stage, potentially affecting the formation and
development of cosmic structures. It also has broader implications for the
overall dynamics and expansion of the Universe. Moreover, the earlier onset
of cosmic acceleration in the modified Kaniadakis cosmology may provide an
explanation for certain observational discrepancies related to the age of
the Universe. By allowing for an earlier transition, this cosmological model
could reconcile the observed age of the Universe with estimates based on
other cosmological parameters. This finding contributes to our understanding
of the Universe's evolution and highlights the importance of considering
alternative cosmological models.

%%%%%%%%%%%%%%%%%%%%%%%%%%%%%%%%%%%%%%%%%%%%%%%%%%%%%%%%%%%%%%%%%%%%%%%%%%%%%%

\section{Summary and Closing remarks \label{Con}}

The exploration of the modified cosmology offers a promising avenue for
addressing fundamental questions in modern cosmology. It not only expands
the horizons of our comprehension regarding the Universe and its evolution
but also introduces novel perspectives and fertile grounds for research in
the fields of cosmology and theoretical physics.

In this paper, we have investigated the implications of the modified
Kaniadakis entropy into cosmological models. Our analysis focuses on
single-component flat Universes during the matter-dominated and
radiation-dominated eras, as well as a multiple-component flat Universe with
pressure-less matter and a cosmological constant/dark energy, revealing
interesting implications for the age and behavior of the Universe.

Interestingly enough, the results demonstrated that in the
modified Kaniadakis cosmology, the predicted age of the Universe
is greater compared to the standard cosmology. This suggests that
the age problem in cosmology could be alleviated by considering
Kaniadakis entropy. Furthermore, we have observed that the scale
factor decreases in the modified Kaniadakis cosmology compared to
the standard cosmology, implying a decrease in the radius of the
Universe. Our studies also reveal that the transition from a
decelerated Universe to an accelerated Universe occurred at higher
redshifts in the modified Kaniadakis cosmology compared to the
standard cosmology. This implies that in the modified Kaniadakis
cosmology, the onset of cosmic acceleration is ahead compared to
the standard cosmology. Furthermore, our analysis reveals that a
Kaniadakis Universe with a single pressure-less matter component
can undergo a transition from early-time decelerated expansion to
late-time accelerated expansion. These findings are intriguing, as
they not only demonstrate the potential impact of incorporating
Kaniadakis entropy into the geometric component of field
equations, but also highlight the significant implications that
arise from such modifications. Specifically, these modifications
can lead to accelerated expansion at later stages, even in the
absence of dark energy, as well as a greater age for the Universe.
Such results provide valuable insights into the dynamics of the
Universe and challenge our current understanding of its evolution.

Future researches in this field could explore the implications of Kaniadakis
entropy in other cosmological scenarios, such as inflationary models or dark
energy models. Additionally, further investigations could consider the
impact of Kaniadakis entropy on other cosmological observable, such as the
cosmic microwave background radiation or the formation of large-scale
structures. Overall, the findings provides valuable insights into the
implications of Kaniadakis entropy in cosmological models and highlights the
potential for addressing age-related issues and understanding the behavior
of the Universe in different cosmological scenarios.
%%%%%%%%%%%%%%%%%%%%%%%%%%%%%%%%%%%%%%%%%%%%%%%%%%%%%%%%%%%%%%%%%%%%%%%%%%%%%%%%%%%%%
\acknowledgments{MKZ thanks Shahid Chamran University of Ahvaz,
Iran for supporting this work under research grant No.
SCU.SP1401.37271. The work of AS is supported by Iran National
Science Foundation (INSF) under grant No. 4022705.
%%%%%%%%%%%%%%%%%%%%%%%%%%%%%%%%%%%%%%%%%%%%%%%%%%%%%%%%%%%%%%%%%%%%%%%%%%%%%%%%%%%%%%%%%

\end{document}